\documentclass[twocolumn,aps,floatfix]{revtex4}
\usepackage{amssymb} \usepackage{graphicx} \usepackage{amsmath}
\usepackage[T1]{fontenc} \usepackage{pstricks} \usepackage{subfigure}

\begin{document}
\title{Thermal solitons as revealed by static structure factor }

\author{Krzysztof Gawryluk,$^{1}$ Miros{\l}aw Brewczyk,$\,^{1,2}$ and Kazimierz Rz\k a\.zewski$\,^2$}

\affiliation{\mbox{$^1$Wydzia{\l} Fizyki, Uniwersytet w
    Bia{\l}ymstoku,  ul. K. Cio{\l}kowskiego 1L, 15-245 Bia{\l}ystok,
    Poland} \\ \mbox{$^2$Center for Theoretical
    Physics PAN, Al. Lotnik\'ow 32/46, 02-668 Warsaw, Poland}  }

\date{\today}

\begin{abstract}
We study, within a framework of the classical fields approximation, the static structure factor of a weakly interacting Bose gas at thermal equilibrium. As in a recent experiment (R. Schley et al., Phys. Rev. Lett. 111, 055301 (2013)), we find that the thermal distribution of phonons in a three-dimensional Bose gas follows the Planck distribution. On the other hand we find a disagreement between the Planck and phonon (calculated just like for the bulk gas) distributions in the case of elongated quasi-one-dimensional systems. We attribute this discrepancy to the existence of spontaneous dark solitons (i.e., thermal solitons as reported in T. Karpiuk et al., Phys. Rev. Lett. 109, 205302 (2012)) in an elongated Bose gas at thermal equilibrium.

\end{abstract}

\maketitle

Low dimensional systems in which particles are restricted in their spatial motion often exhibit peculiar properties. Among them, a quasi-one-dimensional ultracold Bose gas of atoms interacting by contact forces is an ideal example. Its one-dimensional analogue has been studied theoretically by Lieb and Liniger already a long time ago \cite{LiebI}. A surprising outcome stating that such a system possesses two families of elementary excitations was found \cite{LiebII}. At that time only the meaning of the main branch of the excitations was known. In the limit of weak interactions it corresponds to Bogoliubov phonons \cite{Bogoliubov}. However, the understanding of the other branch was not clear, even more, some questions related to the "double counting" of excitations were raised \cite{LiebII}.

By analyzing the zero temperature dispersion curve for solitary waves it was conjectured that the type II branch of elementary excitations represents dark solitons \cite{darksolitons}. This supposition was further confirmed by beyond mean-field studies of quantum dark solitons \cite{darksolCarrUeda} and by exploring the statistical distribution of elementary excitations of weakly interacting Bose gas at thermal equilibrium \cite{Karpiuk12, Karpiuk15}. It was also shown that a particular superposition of the type II eigenstates of the Lieb-Liniger model leads to the density notch in the reduced one-particle density \cite{Sato}. This density initially perfectly overlaps the plot of the density of a dark soliton in the weak coupling limit. Finally, it was demonstrated that a single type II eigenstate reveals a dark soliton during the measurement of particle positions \cite{Sacha}. Even more, calculating the many-body time correlation functions allows to investigate the soliton's dynamics \cite{Syrwid}.

Type II excitations of the Lieb-Liniger model have never been observed even though the elongated weakly interacting Bose gas is currently commonly produced in the experiments \cite{1Dgases}. The direct detection of thermal solitons (i.e., the type II excitations in the weak interaction limit) with the existing technical possibilities is difficult since it requires, first, an access to the $\it in\, situ$ dynamics of the gas and, second, the use of high enough spatial resolution apparatus. Therefore, we propose an indirect way of observing the type II excitations. Some signatures of existence of such excitations can be already seen through the results of experiment of Ref. \cite{Perrin12}, see analysis in \cite{g2}. In another experiment \cite{Meinert}, the type II excitations are probed through the broadening of the dynamical structure factor visible while going from weakly to strongly interacting regime. Here, we demonstrate that comparing the distribution of thermal phonons, obtained via the static structure factor as in \cite{SteinhauerI}, with the Planck distribution for an elongated weakly interacting Bose gas should unambiguously reveal the presence of thermal solitons in the gas. This experimental proposal remains in a direct connection to old experiments with superfluid $^4$He in which the dynamic structure factor was measured to uncover the existence of rotons in the excitation spectrum of liquid helium \cite{helium}.

Below, we are closely following the prescription given in Ref. \cite{SteinhauerI}. To find the distribution of thermal phonons, we calculate the static structure factor according to its definition \cite{PitaevskiiStringari}
\begin{equation}
S({\bf{k}}) = \frac{1}{\langle N\rangle} [\langle |\rho_{\textbf{k}}|^2 \rangle - |\langle \rho_{\textbf{k}} \rangle|^2] \,,
\label{ssf}
\end{equation}
where $\rho_{\textbf{k}}$ is the Fourier transform of the atomic density. The symbol $\langle\, \rangle$ means here the average over grand canonical ensemble. In particular, the quantity $\langle N\rangle$ appearing in Eq. (\ref{ssf}) is an average of total number of atoms over the samples.

Both condensed and thermal atoms have contributions to the static structure factor. In the simplest approximation it is assumed that both these entities contribute independently and therefore the total static structure factor can be written as \cite{SteinhauerI}
\begin{equation}
S({\bf{k}}) = \frac{\langle N_c\rangle}{\langle N\rangle}\, S_c({\bf{k}}) +
              \frac{\langle N_{th}\rangle}{\langle N\rangle}\, S_{th}({\bf{k}})   \,.
\label{ssftotal}
\end{equation}
The static structure factor for the weakly interacting uniform Bose gas at very low temperatures (i.e., much smaller than the critical one) can be calculated within the Bogoliubov approximation \cite{Bogoliubov}. At zero temperature it is given by
\begin{equation}
S_0({\bf{k}}) = \frac{\hbar^2 k^2}{2 m \epsilon(k)}  \,.
\label{ssf0}
\end{equation}
For nonzero temperatures it is \cite{PitaevskiiStringari}
\begin{equation}
S_c({\bf{k}}) =   S_0({\bf{k}})\, \coth\frac{\epsilon(k)}{2 k_B T}  \,
\label{ssfBog}
\end{equation}
and can be put in an equivalent way
\begin{equation}
S_c({\bf{k}}) = (N_{\textbf{k}} + N_{-\textbf{k}} + 1)\,  S_0({\bf{k}})  \,,
\label{ssfBog1}
\end{equation}
where $N_{\textbf{k}}=(\exp(\epsilon(k)/k_B\,T)-1)^{-1}$ is an average number of phonons in a mode with momentum $\hbar \textbf{k}$ and the energy 
\begin{equation}
\epsilon(k) = \left[ \left(\frac{\hbar^2 k^2}{2m}\right)^2 + \frac{\hbar^2 k^2}{m} g\, n \right]^{1/2} \,.
\label{Bogene}
\end{equation}
Here, $n$ means the density of a uniform gas and $g$ is the interaction coupling constant.

On the other hand, the contribution to the static structure factor coming from the thermal cloud can be calculated as \cite{LandauLifshitz,SteinhauerII}
\begin{equation}
S_{th}({\bf{k}}) = 1+\frac{1}{(2\pi)^3\, n} \int n_{\textbf{k}^\prime} 
                          n_{\textbf{k}^\prime + \textbf{k}}  d^{\,3}k^\prime \,
\label{ssfther}
\end{equation}
with atomic populations given by the Bose distribution $n_{\textbf{k}}=(\exp((\hbar^2 k^2/2 m -\mu)/k_B\,T)-1)^{-1}$, where $\mu$ is the chemical potential of a thermal gas. For low temperatures considered in this work, the second term in Eq. (\ref{ssftotal}) can be safely neglected. This is because for temperatures below $0.2 T_c$ the condensate depletion is less than $1\%$. Therefore we have $S({\bf{k}})=S_c({\bf{k}})$.

Our strategy is then as follows. First, we numerically prepare the grand canonical ensemble of classical fields and calculate the static structure factor according to Eq. (\ref{ssf}). Second, since our sample is nonuniform, we must average Eq. (\ref{ssfBog1}) as due to the local density approximation (LDA) \cite{SteinhauerRMP}. Since for low temperatures $S({\bf{k}})=S_c({\bf{k}})$, then from (\ref{ssfBog1}) we have 
\begin{equation}
S({\bf{k}}) = \langle S_c({\bf{k}}) \rangle_{LDA} = \langle (2 N_{\textbf{k}} + 1)\,  S_0({\bf{k}}) \rangle_{LDA}   \,.
\label{Planck}
\end{equation}
To get the average number of phonons in mode $\hbar \textbf{k}$ we decorrelate the right-hand side of Eq. (\ref{Planck}) and find
\begin{equation}
\langle N_{\textbf{k}} \rangle_{LDA} =
\frac{1}{2} \frac{S({\bf{k}})}{\langle S_0({\bf{k}}) \rangle_{LDA}} - \frac{1}{2}  \,.
\label{Planck1}
\end{equation}
Simultaneously, based on the Planck distribution
\begin{equation}
\langle N_{\textbf{k}} \rangle_{LDA} =
\left\langle  \frac{1}{e^{\epsilon(k)/k_B T}-1} \right\rangle_{LDA}  \,.
\label{Planck2}
\end{equation}
Averaging due to the local density approximation means integrating with the atomic density distribution \cite{SteinhauerRMP} $\langle ... \rangle_{LDA}=\int ...\, n(\textbf{r}) d^{\,3}r/ \int n(\textbf{r}) d^{\,3}r$. Our goal is to compare the results obtained with the help of Eqs. (\ref{Planck1}) and (\ref{Planck2}) for three-dimensional as well as for quasi-one-dimensional samples.

Several techniques have been already developed to study degenerate Bose gases at nonzero temperatures, see reviews \cite{review, Proukakis, Blakie} for the description of the methods and their applications. Here, we are following the classical fields approximation (CFA) \cite{CFA,review}. We generate, by a Monte Carlo method using the Metropolis algorithm, the grand canonical ensemble of the classical fields. The need for the grand canonical ensemble is obvious since otherwise $S({\bf{k}=0})=\langle \delta N^2 \rangle / \langle N\rangle$ vanishes whereas $S_c({\bf{k}=0})$, calculated within the Bogoliubov approximation, does not (see Eq. (\ref{ssfBog})). This is because the formula of Eq. (\ref{ssfBog}) is obtained within the formalism not preserving the total number of atoms. The method for generating the grand canonical ensemble closely follows the method developed by us previously for the canonical ensemble \cite{Witkowska}. Now, we have two control parameters: the temperature of the thermostat and the chemical potential of the reservoir of particles. To check the quality of our algorithm we verify the following constraints which must be fulfilled at thermal equilibrium
\begin{eqnarray}
&& \mu = \left( \frac{\partial F}{\partial N} \right)_T     \label{test1} \\ 
&& k_B T = \frac{\langle \delta N^2 \rangle}{\left( \frac{\partial N}{\partial \mu} \right)_T}   \,,
\label{test2}
\end{eqnarray}
where $N$ and $F$ are the number of atoms and the free energy of the system, respectively. For low temperatures the free energy in Eq. (\ref{test1}) can be safely replaced just by the energy of the system.

\begin{figure}[htb] 
\includegraphics[width=8.5cm]{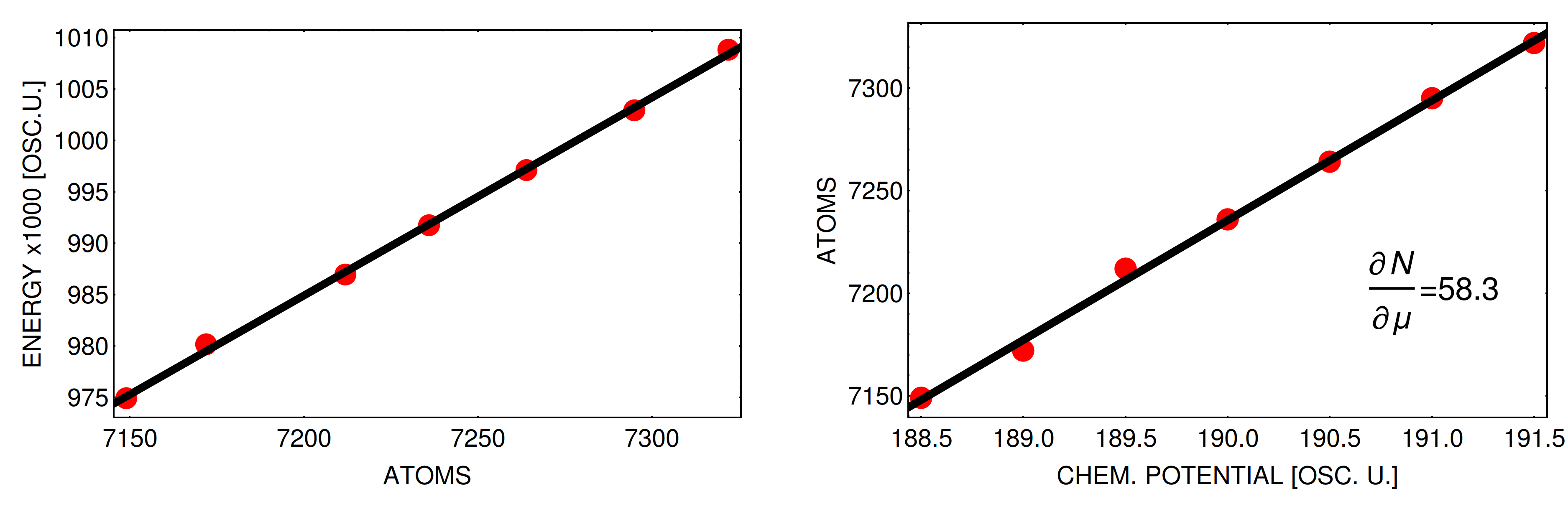}
\caption{(color online).  Left frame: Verification of the condition given by Eq. (\ref{test1}). The figure shows the energy as a function of the number of atoms. The values of the control parameters for the grand canonical ensemble are: $T=326$ and $\mu=190$ in oscillatory units. The linear fit to numerical points gives the value of the chemical potential which equals $192.4$. Right frame: The number of atoms as a function of the chemical potential. Here, the linear fit to numerical data gives the derivative $(\partial N / \partial \mu)_T = 58.3$ and from Eq. (\ref{test2}), knowing that $\langle \delta N^2 \rangle=18950$ we obtain $T=325$. }
\label{quality}
\end{figure}
The result of such a verification is shown in Fig. \ref{quality}. Here, we consider a one-dimensional Bose gas of a few thousands of $^{87}$Rb atoms confined in a harmonic trap with the frequency equal to $2\pi \times 10\,$Hz. Keeping constant thermostat temperature $T=326$ and changing slightly the chemical potential of the reservoir around $\mu=190$ we plot the average energy of the system as a function of the average number of atoms (left frame in Fig. \ref{quality}). Calculating the slope of the linear fit gives us the chemical potential of the system as $192.4$ which is very close to the reservoir's chemical potential. On the other hand, analyzing the average number of atoms as a function of the chemical potential (right frame in Fig. \ref{quality}) we obtain the derivative $(\partial N / \partial \mu)_T$ appearing in the denominator of condition (\ref{test2}). Knowing the variance of the number of particles in the ensemble we calculate the right-hand side of (\ref{test2}) which is $T=325$ -- the value very close to the environment's temperature.

\begin{figure}[htb] 
\includegraphics[width=6.3cm]{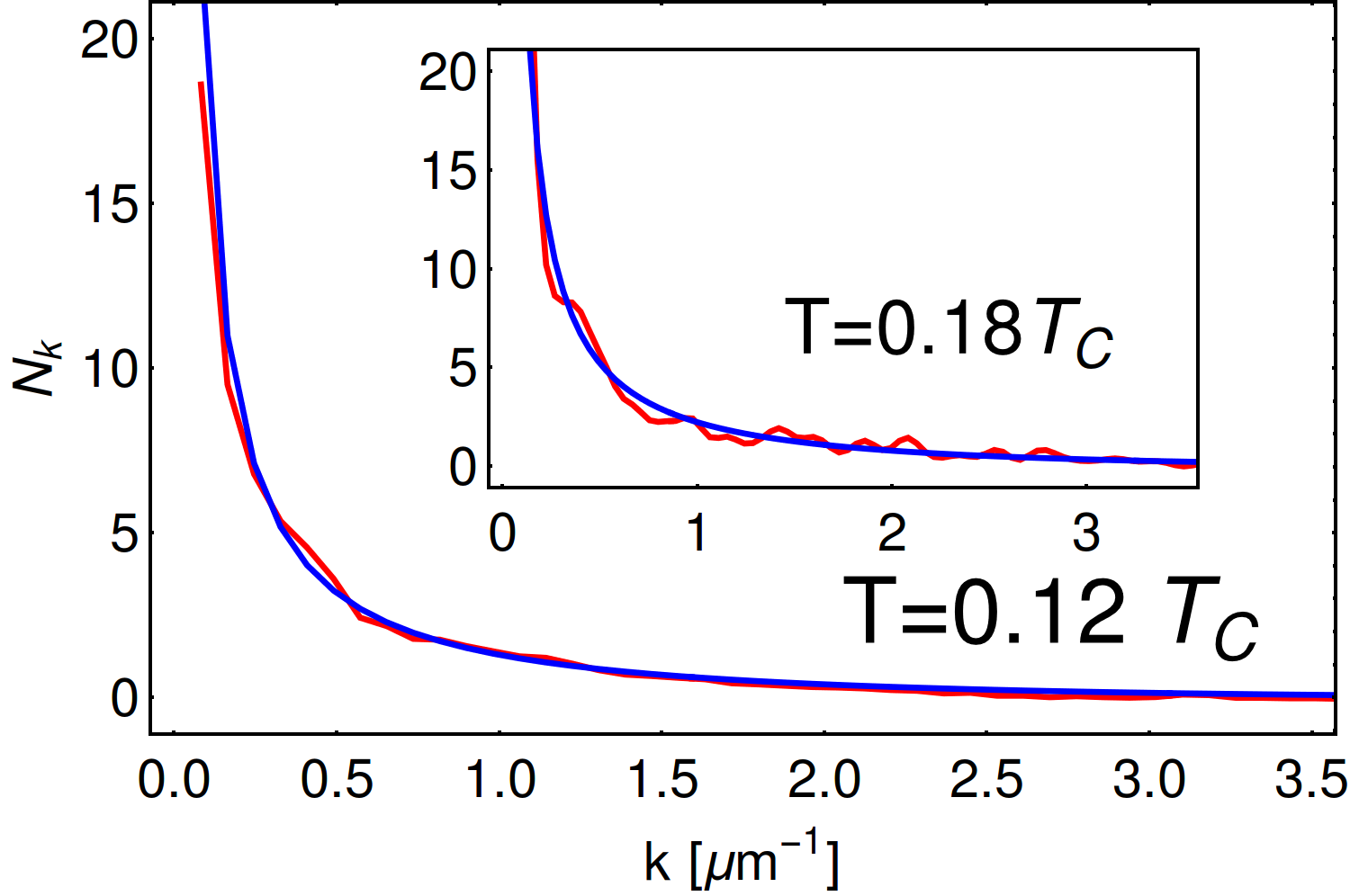}
\caption{(color online). Average number of thermal phonons as a function of momentum in a Bose gas at temperature $T=0.12 T_c$ (main frame) and $T=0.18 T_c$ (inset). The gas is confined in an axially symmetric harmonic trap with radial and axial frequencies $\omega_r=2\pi \times 224\,$Hz and $\omega_z=2\pi \times 26\,$Hz (like in the experiment of Ref. \cite{SteinhauerI}), respectively. An average number of atoms in a trap equals $<N>=72000$. $T_c$ is the critical temperature of an ideal Bose gas with above mentioned parameters. The red curve comes from the analysis of the static structure factor within the classical fields approximation (Eq. (\ref{Planck1})) whereas the blue one is the Planck distribution (Eq. (\ref{Planck2})). In both cases the local density approximation is used. A good agreement between the classical fields approximation result and the Planck distribution is evident.  }
\label{main}
\end{figure}

Now we start to analyze the three-dimensional Bose gas at the parameters studied experimentally in Ref. \cite{SteinhauerI}. We consider an atomic cloud of $^{87}$Rb atoms confined in a harmonic trap with radial and axial frequencies equal to $\omega_r=2\pi \times 224\,$Hz and $\omega_z=2\pi \times 26\,$Hz, respectively. An average number of atoms is $<N>=72000$. Following the prescription described above we calculate the static structure factor and then according to the formula (\ref{Planck1}) we obtain the average number of phonons in each mode (see red curves in Fig. \ref{main}). At the same time we calculate the distribution of thermal phonons based on the Planck law (blue curves in Fig. \ref{main}). For both temperatures considered here both distributions match very well. This agreement is actually expected since the Planck distribution of thermal phonons in a three-dimensional Bose gas was already observed experimentally in \cite{SteinhauerI}.

\begin{figure}[htb] 
\includegraphics[width=6.3cm]{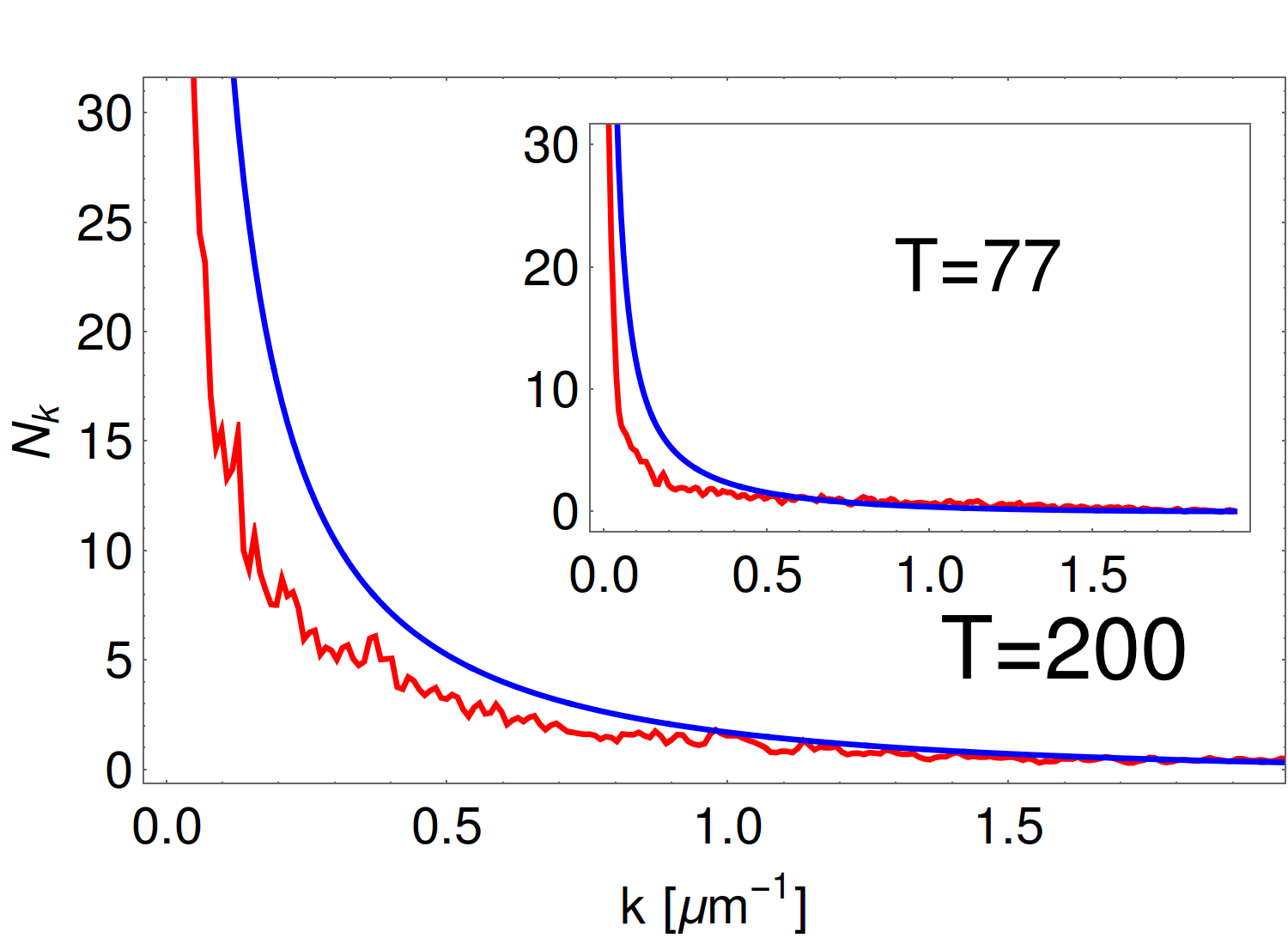} 
\caption{(color online). Average number of thermal phonons as a function of momentum in a thermal Bose gas confined in an elongated axially symmetric harmonic trap with radial and axial frequencies $\omega_r=2\pi \times 113\,$Hz and $\omega_z=2\pi \times 1\,$Hz, respectively. The temperatures are: $T=200$ (main frame) and $T=77$ (inset) in units of $\hbar \omega_z /k_B$. An average number of atoms in a trap equals approximately $<N>=16000$. The red curve comes from the analysis of the static structure factor within the classical fields approximation (Eq. (\ref{Planck1})) whereas the blue one is the Planck distribution (Eq. (\ref{Planck2})). In both cases the local density approximation is used. Here, a disagreement between the classical fields approximation result and the Planck distribution appears for lower momenta.  }
\label{mainq1D}
\end{figure}

Now we raise the question whether the formulas, Eqs. (\ref{Planck1}) and Eq. (\ref{Planck2}), produce the same distributions in the case of a quasi-one-dimensional Bose gas. This is a legitimate question because, as we know \cite{LiebI,LiebII}, a one-dimensional Bose gas has qualitatively new properties with respect to the bulk gases. A one-dimensional system exhibits two branches of elementary excitations, belonging to phonons \cite{Bogoliubov} and to dark solitons \cite{darksolitons,darksolCarrUeda,Karpiuk12,Karpiuk15,Sacha,Syrwid}. The static structure factor calculated within the classical fields approximation via Eq. (\ref{ssf}) obviously includes contributions both from phonons and dark solitons. This is because both these kinds of excitations can be identified within the CFA \cite{phonons,Karpiuk15}. On the other hand, the formula (\ref{Planck2}) considers only phonons. Therefore, for one-dimensional systems there should be a difference between distributions obtained based on (\ref{Planck1}) and (\ref{Planck2}). It is indeed, as shown in Fig. \ref{mainq1D}. Clearly, the disagreement appears for momenta lower than $1\,\mu$m$^{-1}$ what is related to the radial size of the atomic cloud we work with. For phonons at the wavelengths shorter than the radial size of the Bose gas the quasi-one-dimensional character of the system does not matter -- they behave as in a three-dimensional bulk system. Fig. \ref{mainq1D} shows that the discrepancy between the distributions resulted from (\ref{Planck1}) and (\ref{Planck2}) gets larger for higher temperatures. It is expected behavior since at higher temperatures the number of deeper thermal solitons increases \cite{Karpiuk15} causing larger density fluctuations.

To make our analysis conclusive we finally consider the purely one-dimensional Bose gas in a box with periodic boundary conditions. In this case we need no additional simplifying assumptions (neglect of thermal contribution or LDA). We numerically generate the grand canonical ensemble for such a system and calculate the static structure factor according to (\ref{ssf}). We also calculate the density correlation function, $g_2(x)$, as in Ref. \cite{g2} and obtain the static structure factor by taking its Fourier transform $S(k)=n\int dx\, [g_2(x)-1] \exp{(ikx)}$. Results we get in these two ways match each other very well. We compare the static structure factor obtained within the CFA approximation with the one calculated based on the extension of Bogoliubov theory to quasicondensates \cite{Mora03}, valid when $k_B T / (\mu n \xi) \ll  1$, where $\xi=\hbar/\sqrt{m\mu}$ is the healing length and $n$ is the average density. The latter considers only phonons whereas the CFA approach includes both phonons and dark solitons. Hence, the differences between these two approaches are expected.  

\begin{figure}[htb] 
\includegraphics[width=4.5cm]{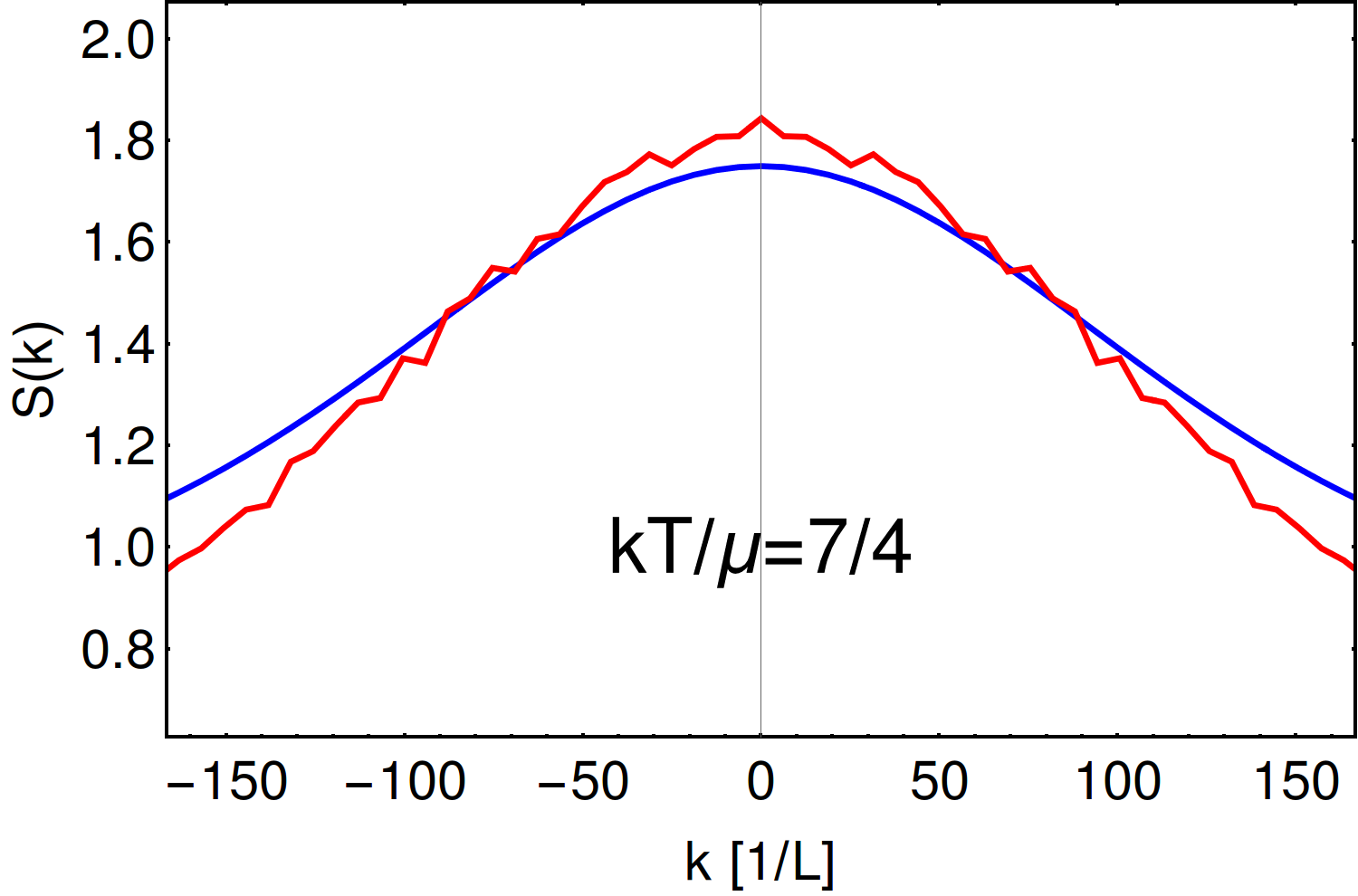}   \hfill
\includegraphics[width=3.8cm]{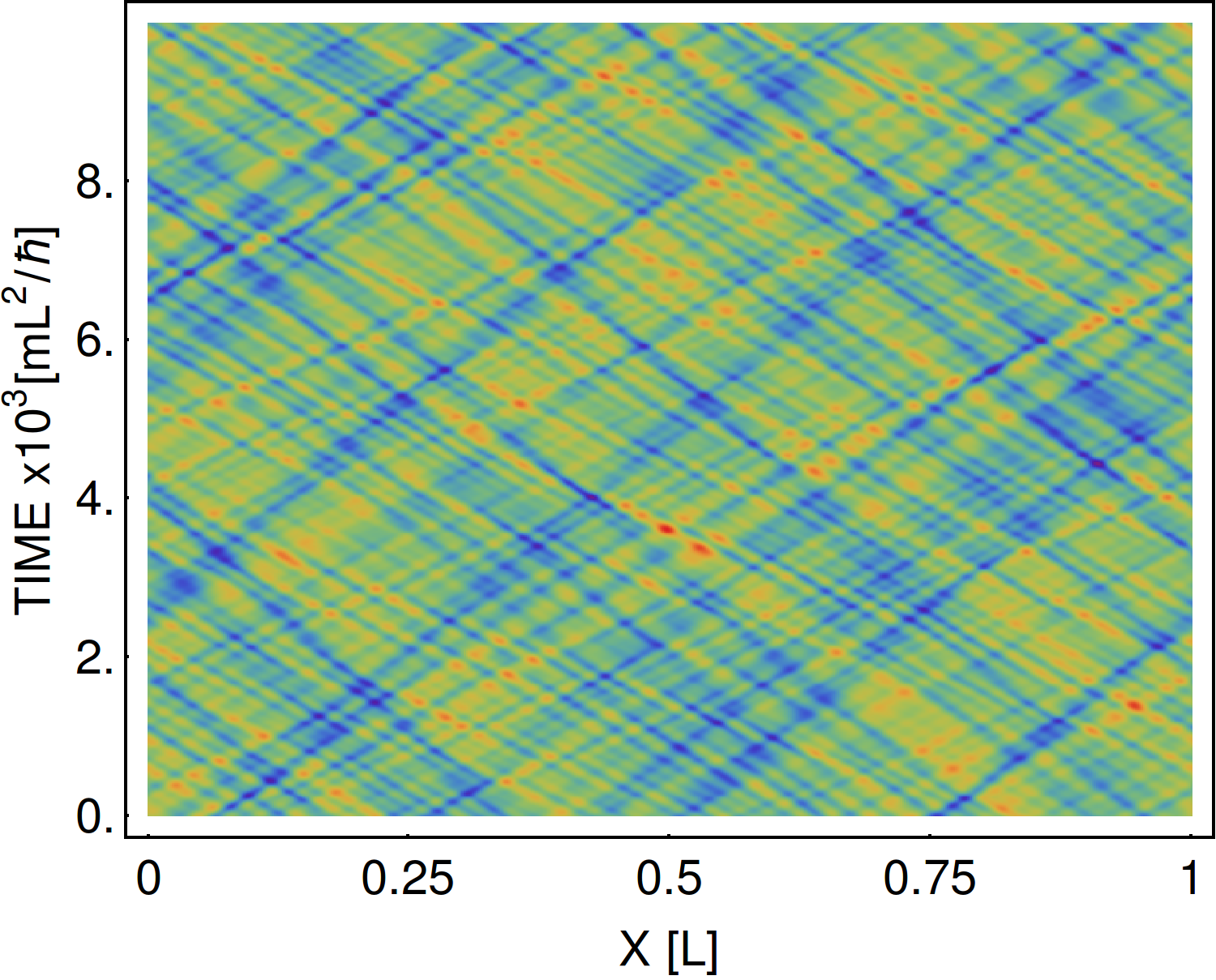}  \\   \vspace{0.4cm}
\includegraphics[width=4.5cm]{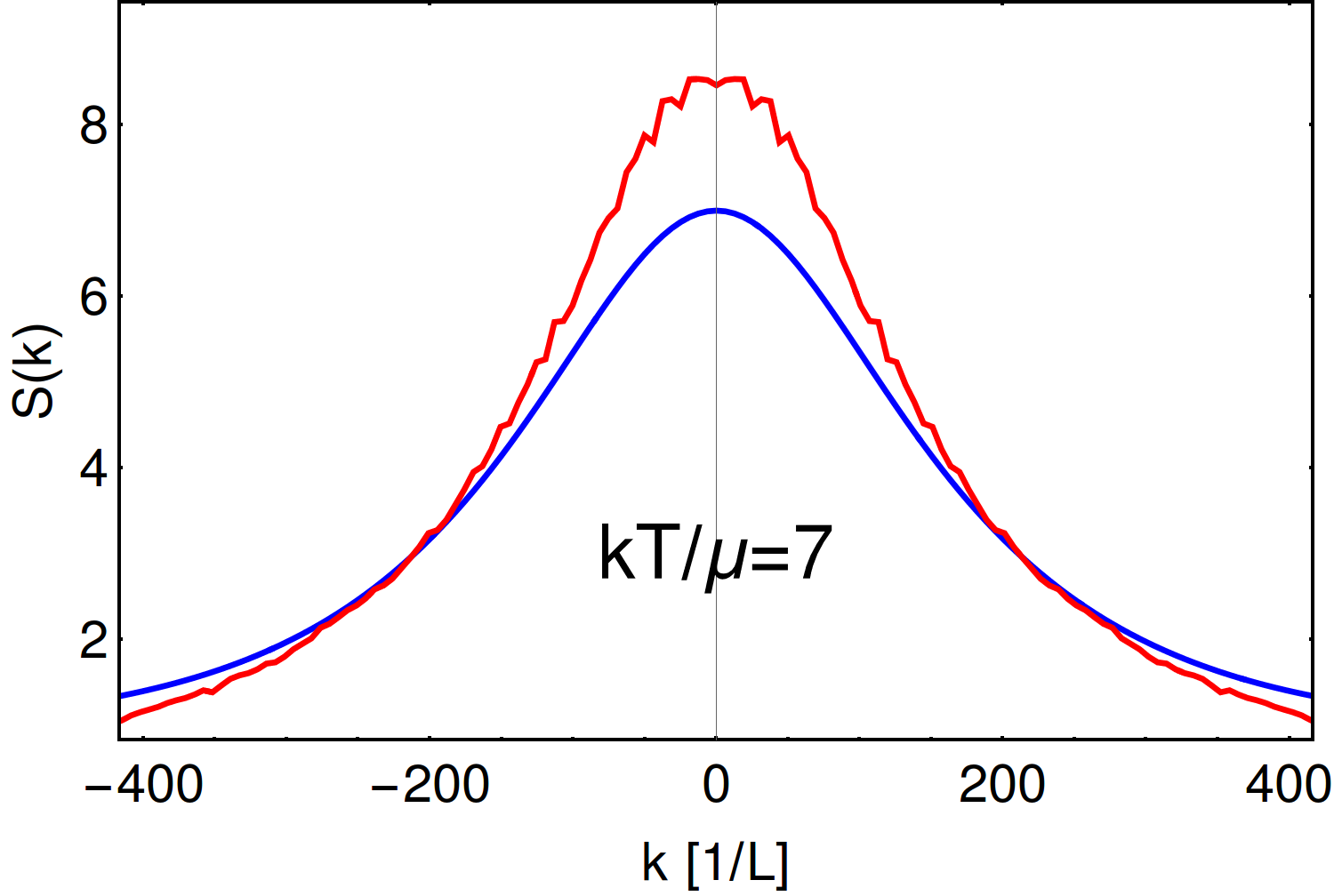}    \hfill
\includegraphics[width=3.8cm]{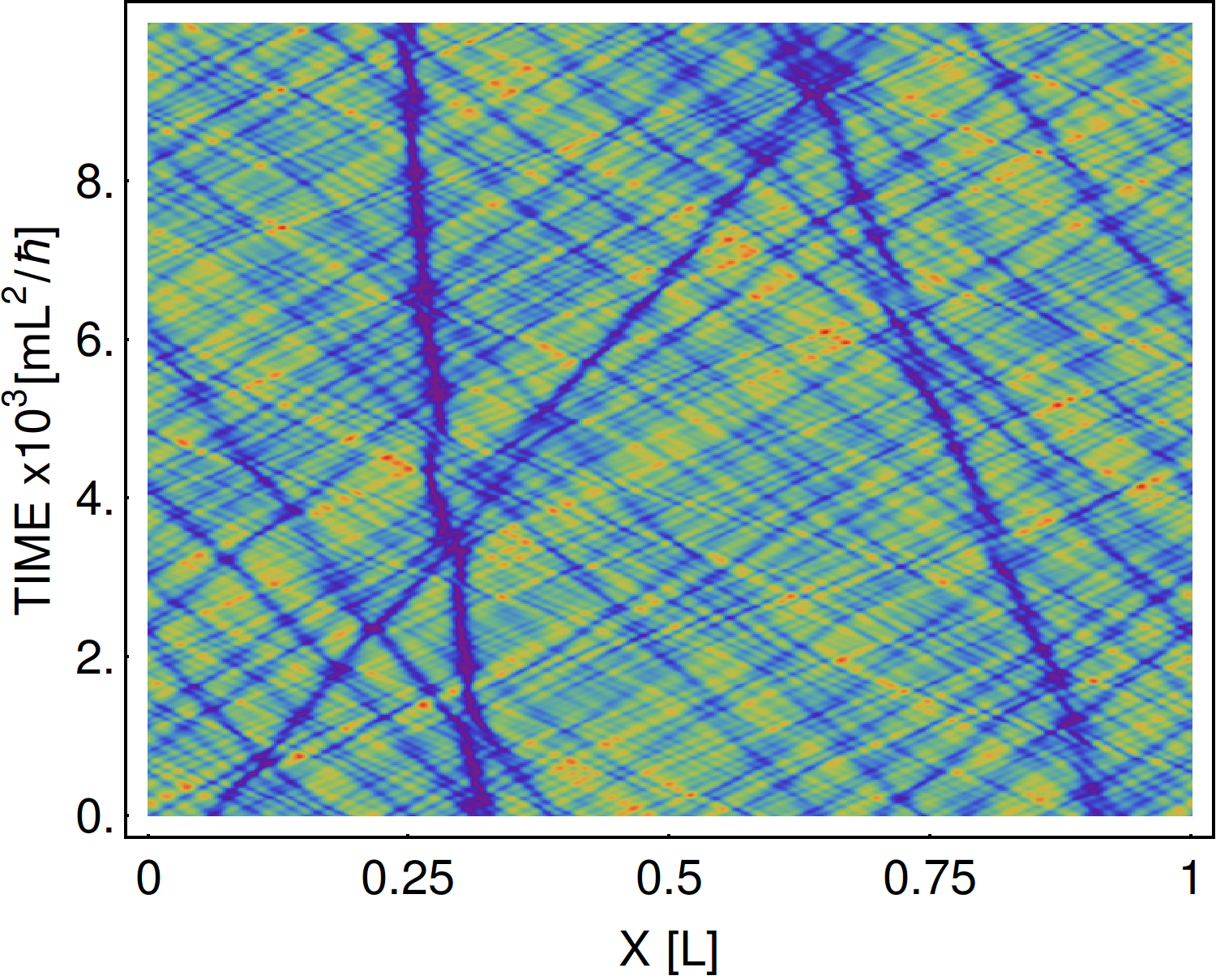} 
\caption{(color online). Left column: Static structure factor $S(k)$ from the CFA approximation (red curves,) and the extension of Bogoliubov theory \cite{Mora03} (blue lines). Right column: Linear density of a 1D Bose gas at equilibrium (from CFA) as a function of time. The upper frames correspond to the temperature $T=1.4\times 10^4\, \hbar^2/mL^2k_B$, where $L$ is the size of the box. The temperature for the lower frames case is four times higher $T=5.6\times 10^4\, \hbar^2/mL^2k_B$. The chemical potential equals $\mu = 8000$ in units of $\hbar^2/mL^2$ for upper and lower case. The average number of atoms is about $<N>=2600$ in both cases. Hence, the dimensionless Lieb-Liniger parameter is $\gamma \approx 0.001$ and the reduced temperature   \cite{Deuar} equals $\tau=0.004$ and $\tau=0.016$ for upper and lower case, respectively } 
\label{box1D}
\end{figure}

We do CFA calculations of the static structure factor in the regime of thermally excited quasicondensate \cite{Deuar} (for based on the Lieb-Liniger model calculations in the high-temperature fermionization or Tonks-Girardeau regions see Ref. \cite{Panfil14}). For lower temperatures the $S(k)$ obtained from the CFA method follows the curve calculated from the phonon only based approach \cite{Mora03} (Fig. \ref{box1D}, left upper frame). Some differences for large momenta are expected and are related to the short wavelength cutoff inherently built in the CFA method \cite{review}. The agreement is expected  since for low temperatures and in the weakly interacting regime only very shallow solitons are excited (see right upper frame in Fig. \ref{box1D} showing the evolution of the linear density of a gas). However, for higher temperatures differences between the CFA and Mora-Castin (phonon only based approach) results appear for lower momenta (Fig. \ref{box1D}, left lower frame). As it is shown in the right lower frame in Fig. \ref{box1D}, deep solitons (seen as almost vertical blue lines) are clearly present in the gas. These dark solitons make the contribution to the static structure factor. For even higher temperatures when the system enters the decoherent quantum regime \cite{Deuar}, the number of thermal solitons significantly increases.

In summary, we have studied the weakly interacting Bose gas at thermal equilibrium, confined in both a three-dimensional and quasi-one-dimensional harmonic traps. We were particularly focused on a distribution of thermal phonons. We found that for three-dimensional samples the distribution of thermal phonons matches the Planck distribution. On the other hand, for quasi-one-dimensional geometries there appears a discrepancy when a 3D proceudre is directly applied, which we attribute to the existence of, in addition to phonons, other kind of elementary excitations, the type II excitations. These excitations, i.e., dark solitons, contribute as well to the static structure factor of an elongated Bose gas. Therefore, the disagreement between the distributions coming from the Bogoliubov approximation (which includes only phonons) and CFA (including both phonons and dark solitons) is expected. This disagreement is an indirect proof of existence of the type II excitations (i.e., thermal solitons). Hence, a measurement of the static structure factor in an elongated weakly interacting Bose gas should unequivocally show the presence of thermal solitons in the gas.

\acknowledgments  
We thank J. Steinhauer for inspiring us to undertake this study and M. Gajda for helpful discussions. 
The work was supported by the National Science Center (Poland) Grant No. DEC-2012/04/A/ST2/00090.

\end{document}